\documentclass{article}

\usepackage[english]{babel}
\usepackage{multirow}

\usepackage[letterpaper,top=2cm,bottom=2cm,left=3cm,right=3cm,marginparwidth=1.75cm]{geometry}
\usepackage{subcaption}
\usepackage{amsfonts}
\usepackage{amsmath}
\usepackage{graphicx}
\usepackage[colorlinks=true, allcolors=blue]{hyperref}
\usepackage{tabularx}
\usepackage{makecell}
\usepackage{array}

\title{Crop Disease Classification using Support Vector Machines with Green Chromatic Coordinate (GCC) and Attention based feature extraction for IoT based Smart Agricultural Applications}
\author{Shashwat Jha\textsuperscript{1*},Vishvaditya Luhach\textsuperscript{2*}, Gauri Shanker Gupta\textsuperscript{1}, Beependra Singh\textsuperscript{3}}
\date{}
\begin{document}
\maketitle

\begin{abstract}
Crops hold paramount significance as they serve as the primary provider of energy, nutrition, and medicinal benefits for the human population. Plant diseases, however, can negatively affect leaves during agricultural cultivation, resulting in significant losses in crop output and economic value. Therefore, it is crucial for farmers to identify crop diseases. However, this method frequently necessitates hard work, a lot of planning, and in-depth familiarity with plant pathogens. Given these numerous obstacles, it is essential to provide solutions that can easily interface with mobile and IoT devices so that our farmers can guarantee the best possible crop development. Various machine learning (ML) as well as deep learning (DL) algorithms have been created and studied for the identification of plant disease detection, yielding substantial and promising results. This article presents a novel classification method that builds on prior work by utilising attention-based feature extraction, RGB channel-based chromatic analysis, Support Vector Machines (SVM) for improved performance, and the ability to integrate with mobile applications and IoT devices after quantization of information. Several disease classification algorithms were compared with the suggested model, and it was discovered that, in terms of accuracy, Vision Transformer-based feature extraction and additional Green Chromatic Coordinate feature with SVM classification achieved an accuracy of (GCCViT-SVM) - 99.69\%, whereas after quantization for IoT device integration achieved an accuracy of - 97.41\% while almost reducing 4x in size. Our findings have profound implications because they have the potential to transform how farmers identify crop illnesses with precise and fast information, thereby preserving agricultural output and ensuring food security. 
\end{abstract}


\textbf{\underline{Affiliations}}\\
\textsuperscript{1} Department of Electrical and Electronics Engineering, Birla Institute of Technology, Mesra, Ranchi 835215, Jharkhand, India\\
\textsuperscript{2} Department of Computer Science and Engineering, Maharaja Surajmal Institute of Technology, New Delhi - 110058, India\\
\textsuperscript{3} Department of Remote Sensing, Birla Institute of Technology (BIT), Mesra-835215, Ranchi. Jharkhand, India\\
\textsuperscript{*} Authors contributed equally \\

\section{Introduction}

Plant diseases and unfavourable soil conditions have become more common in recent years as a result of factors including globalisation, commerce, and climate change (Kamilaris and Prenafeta-Boldú, 2018; Iqbal et al., 2018)\cite{R1,R2}. This has caused a pandemic-like scenario in many nations, endangering the health of crops and the ability of people to acquire enough food and nourishment (Mohanty et al., 2016)\cite{R3}. Since various parasitic organisms, including bacteria, fungus, viruses, nematodes, and other plants, can cause illness, experts must prioritize protecting agricultural plants (Barbedo, 2016\cite{R4}). Accurate diagnosis is necessary in order to handle the concerns raised properly. Several diagnosis techniques include conventional procedures incorporating neighbourhood plant nurseries and agricultural organisations. These techniques could be constrained, though, by human error and the difficulties of reaching plants across wide distances. As an alternative, utilising software to use machine learning and deep learning approaches can improve the accuracy of diagnosing damaged plants. New plant disease categorization systems may now be implemented into a wide range of devices thanks to the development of new embedded devices with high computing power, cameras, and integration modules. Farmers may submit field photographs for analysis, diagnosis, and action planning using this method, which also makes it easier to perform precise diagnostic tests and preserve resources.\\
On the other hand, Deep neural networks, machine learning, and sensor-sourced soil data have recently emerged as very effective methods for continually monitoring plant health and spotting disease early warning indications (Pradhan et al., 2022)\cite{R5}. A neural network analyses a picture of a sick plant as input and uses the image's processing to provide information about the connection between the crop and the illness. The optimisation of the network topology, node functions, and edge weights is required to build a deep network that properly translates the input to the intended output. The method of training deep neural networks involves changing network parameters to enhance the mapping. Transformer-based models have excelled in completing this difficult task (Jha et al., 2022)\cite{R6}. Notably, major developments in theory and real-world applications have significantly improved the performance of deep neural networks (Bengio and LeCun 2007; Schmidhuber, 2015)\cite{R7,R8}. The use of artificial intelligence (AI) to create a variety of intelligent solutions for practical applications has gained increasing attention, and results have been encouraging. However, most of the work done thus far often falls short in terms of real-time crop assessment compatibility and compromises either classification performance or a huge memory footprint.\\ 

Moreover, numerous methods, including machine learning models, CNN models, CNN with attention mechanisms, and CNN with Vision Transformer (ViT), have been investigated in recent research on the categorization of plant diseases. To extract pertinent leaf characteristics for machine learning-based classifiers for plant diseases, considerable preprocessing is frequently necessary. However, these models could show decreased accuracy because of the limits of machine learning approaches. For instance, a study suggested a machine learning model built on the PlantVillage dataset using LGBM, although the performance measures were only 94\% or less effective (Tabbakh and Barpanda, 2022)\cite{R9}.\\
Since, Images of plant diseases have been successfully classified using CNN models. It is difficult to encode the direction and location of diseased sections within leaves due to their design, which largely records pixel connections. Therefore, these models' performance measures could not be particularly near to 1. On the PlantVillage dataset, CNN models were trained in several experiments, but the performance metrics fell short of 99\% (Shijie et al., 2017; Jasim and Al-Tuwaijar, 2020; Mohana et al., 2021; Sachdeva et al., 2021)\cite{R11,R12,R13,R14}. The performance measures also fell short of 99\% in tests employing other CNN types on additional datasets (Lin et al., 2019; Kukreja, and Kumar, 2021; Haider et al., 2021; Suri et al., 2023)\cite{R15,R16,R17,R18}. Researchers have looked into merging CNNs or pre-trained CNN models with attention processes to increase accuracy.\\ 
Additionally, the attention mechanism functions as an adaptive filter, assigning weights on a dynamic basis based on the pixel composition. For instance, one research boosted performance by almost 98\% when DenseNet was combined with an attention mechanism (Chen et al., 2021a)\cite{R19}. In another investigation, an attention mechanism-equipped pre-trained MobileNet-V2 model obtained a performance measure of almost 97\% (Chen et al., 2021b)\cite{R20}. Pre-trained CNN models in conjunction with ViT have been shown to perform the best in classifying plant diseases among the various methods. By combining the advantages of CNNs with ViT, this strategy produces metrics that are very near to 1. On the PlantVillage dataset, a research trained the CNN model used by ViT, attaining performance metrics of over 99\% (Thakur et al., 2021)\cite{R21}.\\

This research reveals how diverse categorization methods have advanced our understanding of plant diseases, with attention processes and the incorporation of ViT showing promise for improving accuracy. However, it is evident that the majority of research either focuses on a single task or has a limited scope in terms of the categories of classes within a dataset. At the same time, models primarily use large architectures of convolution-based methods or transfer learning, which demand significant resources and are afflicted with unwanted transfer bias biased due to the nature of the pre-trained dataset, whereas the effective methods have high complexity. The methods and flaws of the suggested study are shown in \textbf{Table 1} This study intends to present a solution for precision agriculture leveraging Vision Transformer for image feature extraction and using RGB Chromatic coordinates (GCC) calculation as a feature along with SVM based classification trained on PlantVillage dataset (Hughes and Salathé, 2015) while ensuring optimization for IoT and device applications using post-training quantization techniques.\\

\begin{table}[h]
\caption{Analysis of existing literature}
\centering
\renewcommand{\arraystretch}{1.5} 
\begin{tabularx}{\textwidth}{XXX}
\hline
\multicolumn{1}{c}{\makecell{Referred studies}} & \multicolumn{1}{c}{\makecell{Approach}} & \multicolumn{1}{c}{\makecell{Shortcomings}} \\
\hline\hline
Tabbakh et al.,2022\cite{R9} & Discussion of performance of Ml algorithms using GLCM and wavelet based statistical features & Reduced accuracy due to completely ML based approach \\

Feng et al.,2019\cite{R10} & Leveraging attention and residual networks- lightweight architecture & Focusing on a narrow plant category and reduced accuracy due to tradeoff on memory footprint and performance  \\

Shijie et al.,2017	\cite{R11}& Utilizes image expansion for preprocessing and transfer learning with pre-trained weights & No discussions about integration for real time application.
Increased complexity, unwanted bias transfer and loss of task specifity due to transfer learning \\

Kukreja et al.,2021\cite{R16} & Leverages matrix based convolutional techniques & Focusing on a single plant/crop category, no integration for real time application discussed. Reduced accuracy due to limitations of purely CNN based approach \\

Haider et al.,2021\cite{R17} & Deep convolutional neural networks used & Focusing on a single plant/crop category and High computational cost due to large architecture\\

Thakur et al.,2021\cite{R21} & Utilizes attention mechanism of Vision transformer with CNN & No discussions on optimization of model for device applications. Very high computational resource requirement \\
   
Tabbakh et al.,2023\cite{R22} & Hybrid implementation of attention based transformer with transfer learning models using pre-trained weights & Focusing on a specific plant/crop category, no integration for real time application discussed. Unwanted bias transfer and loss of task specify. Large computational memory footprint \\
\hline
\end{tabularx}
\end{table}

This paper presents a disease classification model based on Vision Transformer, Chromatic indices and SVM which can be deployed on a local IOT devices for disease identification of crops. The major contributions of this paper are noted as follows: (i) Utilizing hybrid attention based – Vision Transformer (ViT) with Support Vector Machine (SVM) for disease classification, (ii) Leveraging Channel based analysis of Images to calculate GCC and integrate it as a feature for classification and (iii) Implemented post-training quantization and conversion to package the models for deployment on IoT, Web and Smartphone based applications.\\
The subsequent sections of this article are organized such that Section 2 presents the methodology proposed in this study, outlining the approach taken to address the challenges of plant disease detection. It also includes an explanation of various machine learning (ML) and deep learning (DL) methods utilized. Section 3 focuses on the details of the experimental setup, providing information on the datasets, tools, and parameters used in the evaluation of the proposed methodology, whereas in Section 4, the experimental results are presented along with a comprehensive discussion of the findings. This section analyzes and interprets the outcomes of the experiments with the accuracy and effectiveness of the proposed approach. Finally, Section 5 concludes the paper, summarizing the key findings and contributions of the study. 

\section{Materials and Methods}

For disease classification, the proposed method has three steps, initially, the attention mechanism in the Vision transformer is leveraged to extract features, including positional features. Then, each image in the dataset is split into RGB channel, and GCC is calculated after which Support Vector Machine is utilized to classify the images. Apart from the proposed GCCViT-SVM approach, the performance of some of the popular pre-trained transfer learning models, namely – VGG-16 and InceptionV3 were compared. The methodology of both, the proposed as well as the Transfer learning approaches is discussed in the following sections. \textbf{Fig.1} gives an overview of the proposed GCCViT-SVM.
\\
\begin{figure}[h]
  \centering
  \includegraphics[width=0.75\textwidth]{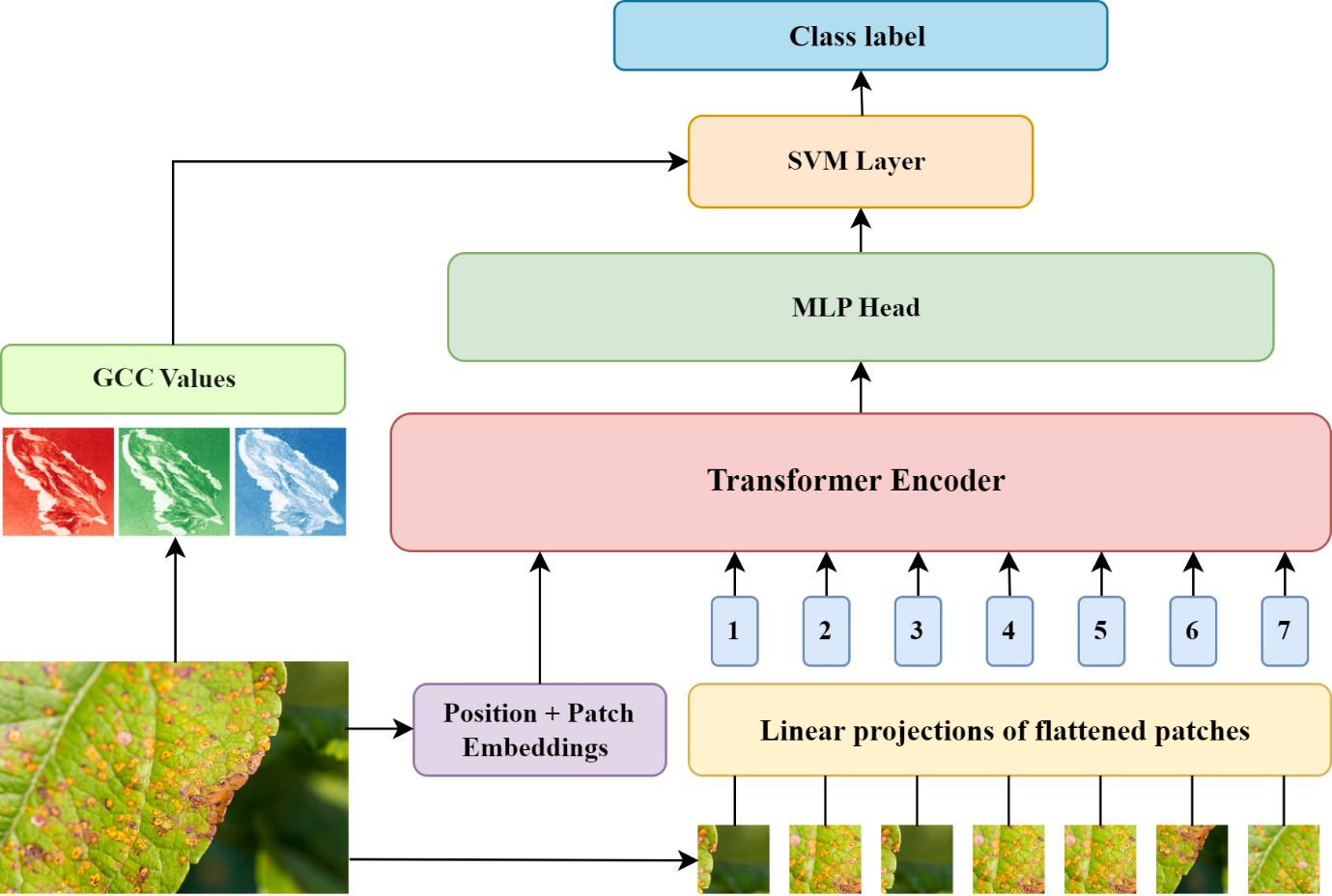}
  \caption{Architecture of GCCViT-SVM}
\end{figure}

\subsection{Visual Feature extraction}
The vision transformer turns the images into flattened patches and converts them into patch embeddings. Along with the patch embeddings the transformer also incorporates standard 1-dimensional positional embeddings to retain the positional information of the patches. An extra learnable class embeddings is also attached before turning it into a 1- dimensional vector which is fed into the transformer encoder \textbf{Fig. 2(a)} illustrates the architecture of transformer encoder which consists of 3 main components as discussed in the following subsections
\subsubsection{Multi Head Self-Attention layer (MSA)}
The multi head self-attention layer is used to calculate the attention using the standard QKV self- attention (Vaswani et al., 2017)\cite{R23}. The input features X are converted into Query (Q), Key (K) and Value (V) matrices which are used to compute self-attention.
\\
\begin{figure}[h]
  \centering
  \begin{subfigure}{0.15\textwidth}
    \centering
    \includegraphics[width=\textwidth]{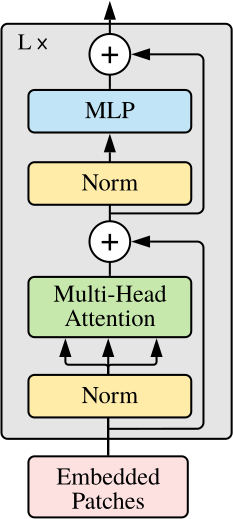} 
    \caption{}
  \end{subfigure}
  \begin{subfigure}{0.42\textwidth}
    \centering
    \includegraphics[width=\textwidth]{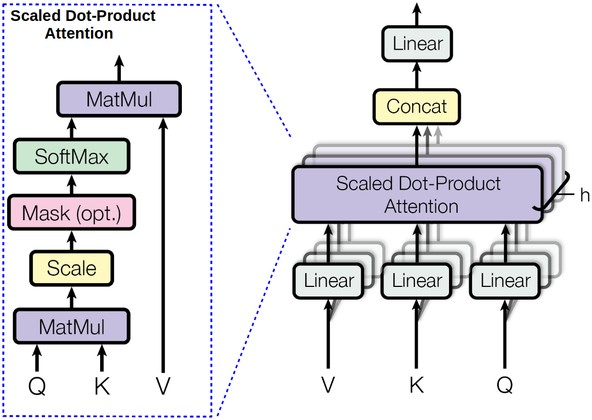} 
    \caption{}
  \end{subfigure}
  \caption{(a) Transformer encoder architecture \ (b) Multi-head self attention}
\end{figure}

 Every element in the input feature $z \in \mathbb{R}^{N \times D}$, a weighted sum is calculated over every value of V. The transformer encoder calculates attention weights as 
\begin{align}
\text{Attention }(Q, K, V) = \text{softmax}\left(\frac{QK^T}{\sqrt{d_k}}\right)V\
\end{align}
Self-attention for the input feature z is calculated using 
\begin{align}
\text{SA}(z) = AV \
\end{align}
Multi head self-attention (MSA) is obtained by running multiple self-attention (SA) operations shown in (2) in parallel and their projected output is concatenated. \textbf{Fig.2(b)} gives an overview of the MSA layer.

\subsubsection{Multi-Layer Perceptron (MLP)}
A multi-layer perceptron refers to a feed-forward artificial neural network. The Vision Transformer MLP block has two layers which consists of Gaussian Error Linear Units (GELU) non-linearity.

\subsubsection{Layer Normalization (LN)}
Layer normalization, according to Ba et al. (2016)\cite{R24} is applied before every block. It is primarily used so as not to introduce new dependencies across two training samples. The transformer encoder architecture alternates MSA and MLP blocks, with LN being applied before every block and a residual connection being established after every block. 

\subsubsection{Green Chromatic Coordinate (GCC)}
Vegetation Indices are important in crop mapping using satellite and PhenoCam data because they provide quantitative assessments of various vegetation features that allow researchers to monitor, estimate, and analyze crop health and growth (Reid et al., 2016; Misra et al., 2020; Grey and Ewers, 2021)\cite{R31,R32,R35}. The Normalised Difference Vegetation Index (NDVI) is a popular remote sensing metric. It can provide data on crop physiological health. Crop mapping indices such as the Normalised Difference Vegetation Index (NDVI), Enhanced Vegetation Index (EVI), and Normalised Difference Water Index (NDWI), as well as their specifications, are used in crop management. Green chromatic coordinates (GCC) are the most often used Phenocam-derived indices from RGB images.\\
Since, most of these images are generated from PhenoCams, which collect images of vegetation throughout time. However, images are collected through UAV and Camera area and captured through mobile phones and balloon stores as RGB. GCC vegetation indices ,used to measure phenological changes in agriculture and vegetation since are comparable and outperformed NDVI (Nijland et al., 2014; Reid et al., 2016)\cite{R34,R35}, are used to track agricultural and vegetation phenological variations. Indices are used for analyzing images from satellite and PhenoCam, assisting researchers and farmers in tracking crop development, detecting stress and disease, estimating crop output, and making intelligent irrigation decisions, fertilization, and pest control decisions (Aydin et al., 2017)\cite{R33}. Moreover, different indices could be more suited for certain crops based on climatic conditions and research objectives.\\ 
Therefore, we use the Plant Village dataset, which contains 61486 images, to generate the green chromatic coordinate (GCC) for classifying Plant diseases. Indices are important in crop mapping with satellite and PhenoCam data because they give quantitative measurements of many vegetation properties, allowing researchers to monitor and analyse crop health and growth. The Green Chromatic Coordinate (GCC), derived from graph theory (Smith, and Smith, 2018)\cite{R28}, has been widely used in agricultural remote sensing. In graph theory, the chromatic index of a graph defines the smallest number of colours required to colour its edges without sharing colours with neighbouring edges. All photos in this work are divided into RGB channels, and the index is determined as
\begin{align}
GCC = \frac{G_{DN}}{{R_{DN} + B_{DN} + G_{DN}}} \
\end{align}
Where, DN is the digital number that corresponds with the green (G), red (R) and blue (B) channels.
\subsection{Classification}
One of the most popular machine learning approach for classification and regression tasks is Support Vector Machines (SVM). This method seeks to identify the best decision threshold that optimizes the margin between classes (Cortes and Vapnik, 1995)\cite{R25}. The kernel approach for non-linear borders is used by SVM to find the support vectors, build a hyper plane, and identify the boundaries. SVM is dependable, manages high-dimensional data, and guards against over fitting. The extracted reduced features from the ViT along with GCC values were further fed into a dense layer with 64 neurons, which was further connected to the final dense layer with 39 neurons (equal to the number of classes) for classification. L2 regularization was applied to this layer to implement the functioning of an SVM. This layer is further optimizing the class boundaries of the features obtained from the Vision Transformer backbone.

\subsection{Quantization and size reduction for device compatibility}
Post-training quantization was utilized to optimize the proposed model for use on Mobile and IoT devices. Post-Training dynamic range quantization was leveraged to transform the initial 32-bit floating-point model into a 8-bit quantized model using Tensorflow Lite. A balance between reduction of large memory footprint for the model while maintaining the high level performance necessary for this application was achieved by quantizing both weights and activations which makes the approach deployable on devices so as to create a dependable solution for disease classification. \textbf{Fig.3} depicts the overall proposed scheme.
\begin{figure}[h]
  \centering
  \includegraphics[width=0.7\textwidth]{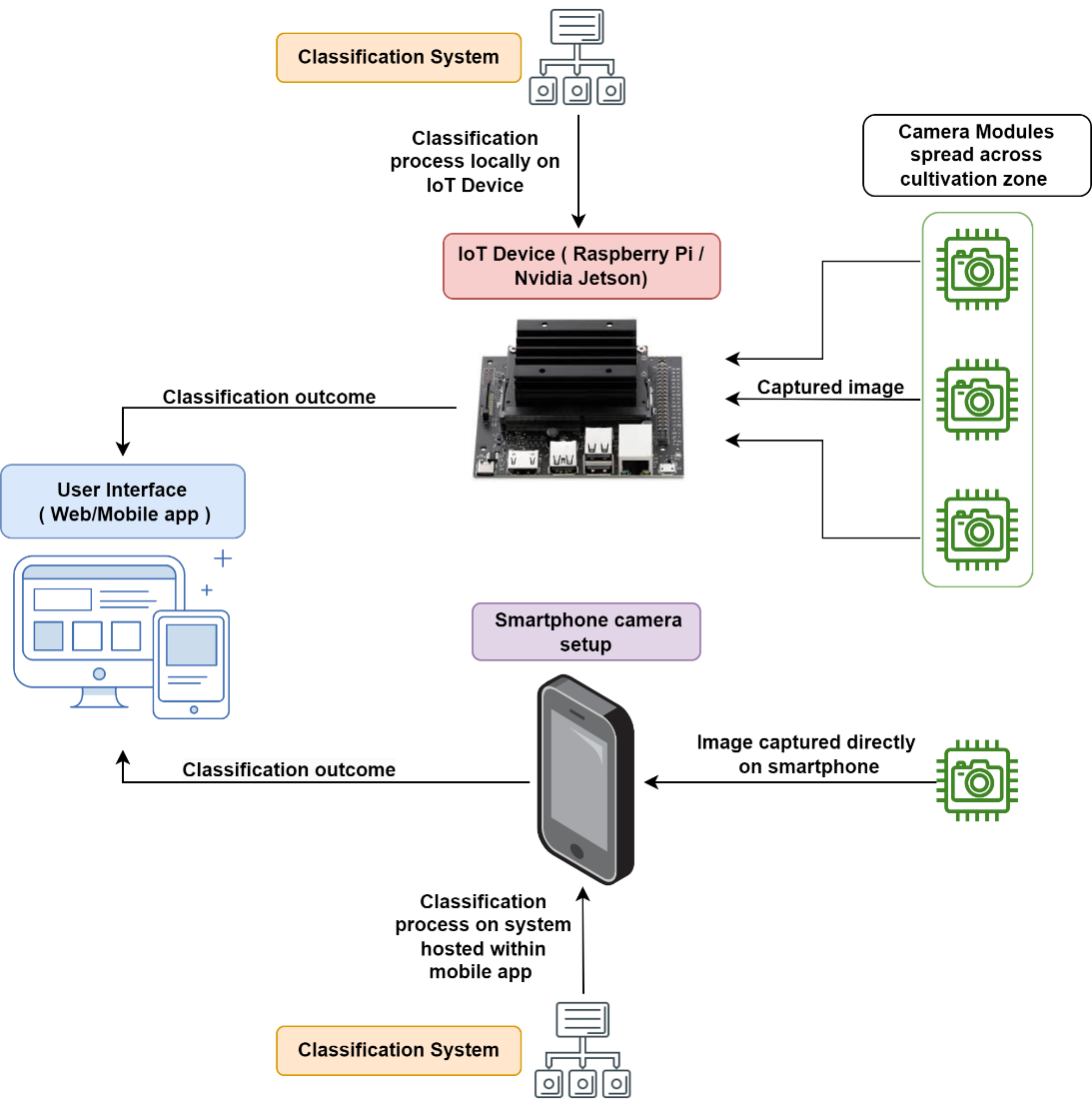}
  \caption{Proposed scheme of solution}
\end{figure}

\subsection{Methods for comparison: Transfer Learning}
For comparing the performance of our model, traditional transfer learning approaches were applied for image classification to the dataset. Transfer learning involves training the earlier layers of a neural network on a different dataset, preferably with huge amount of data, and then training only the last few layers of the neural network on the target dataset. This saves time spent for the training of the model as well as overcomes the issue of not having large enough datasets. Performance of two popular architectures,VGG-16 and InceptionV3 were compared for disease classification

\subsubsection{VGG 16}
VGG-16 is an implementation of the very deep convolutional neural network introduced by Simonyan, and Zisserman, (2014)\cite{R26}. It is a relatively simple convolution-based architecture consisting of 13 convolutional layers and 4 ‘MaxPool’ layers in between. These are followed by 3 fully connected layers and a final layer for classification.

\subsubsection{InceptionV3}
InceptionV3 is also a convolution based neural network from the inception family. It is an improvement upon the existing InceptionV2. It introduces the use of Label Smoothing, 7x7 factorized convolutions and uses an auxiliary classifier which helps to send the label information in the lower levels of the neural network (Szegedy et al., 2016)\cite{R27}. It also utilizes batch normalization in the layers present in the side-head. At the time of its introduction, it performed better than the other state-of-the-art models on standardized datasets.

\subsection{Experimental Analysis}
All models used in this work were trained using GPU acceleration. All the studies were conducted in the Windows 10 operating system running on an AMD Ryzen 3600 CPU @ 3.60 GHz and 16GB RAM along with a NVIDIA 1660ti GPU having 6GB of VRAM. All the models were designed using TensorFlow 2.8.0 framework and the Keras API. For disease classification the plantvillage dataset was utilized (Hughes, and Salathé, 2015)\cite{R29}. In this study, an augmented version of this dataset was utilized, which applies image augmentations such as random clipping, rotation, etc., to increase the total number of images and additional images of backgrounds without any leaves to increase the model's efficiency in real world applications. 
The final dataset used consists of a total of 61,486 images divided across 39 classes. The division of images across classes is not even, with some classes consisting of more than 5000 images whereas others consist of no less than 1000 images. The full details of the classes and their samples is given in \textbf{Table 2}.

\begin{table}[h]
\caption{Analysis of existing literature}
\centering
\renewcommand{\arraystretch}{1.2} 
\small
\begin{tabularx} {\textwidth}{cccc}
\hline
\ Class & Samples & Class & Samples \\
\hline\hline
\ Apple Scab & 1000 & Grape Esca (Black Measles) & 1383 \\
\ Apple Black Rot & 1000 & Bell pepper healthy & 1478\\
\ Apple cedar apple rust & 1000 & Potato Early blight & 1000\\
\ Apple healthy & 1645 & Potato healthy & 1000\\
\ Background without leaves & 1143 & Tomato Septoria leaf spot & 1771\\
\ Blueberry healthy & 1502 & Raspberry healthy & 1000 \\
\ Cherry healthy & 1000 & Soybean healthy & 5090 \\
\ Cherry Powdery mildew & 1052 & Tomato Yellow Leaf Curl Virus & 5357\\
\ Corn Cercospora leaf spot & 1000 & Strawberry healthy & 1000 \\
\ Tomato Leaf Mold & 1000 & Peach Bacterial spot & 2297 \\
\ Corn healthy & 1162 & Tomato Bacterial spot & 2127 \\
  \ Corn Northern Leaf Blight & 1000 & Tomato Early blight & 1000 \\
  \ Grape Black rot & 1180 & Tomato healthy & 1591 \\
  \ Bell pepper Bacterial spot & 1000 & Strawberry Leaf scorch & 1109\\
  \ Grape healthy & 1000 & Orange Haunglongbing (Citrus greening) & 5507 \\
  \ Grape Leaf blight (Isariopsis Leaf Spot) & 1076 & Potato Late blight & 1000\\
  \ Corn Common rust & 1192 & Squash Powdery mildew & 1835\\
   \ Tomato Late blight & 1909 & Tomato Target Spot & 1404\\
   \ Peach healthy & 1000 & Tomato mosaic virus & 1000\\
\ Tomato Spider mites & 1676 \\
\hline
\end{tabularx}
\end{table}

\textbf{Fig. 4} shows a sample of some of the images present in the dataset. Stratified split was utilized to split data into sets for training purposes such that distribution of images across classes in the same ratio to ensure robustness and enhanced training proficiency of the model.

\begin{figure}[h]
  \centering
  \includegraphics[width=0.4\textwidth]{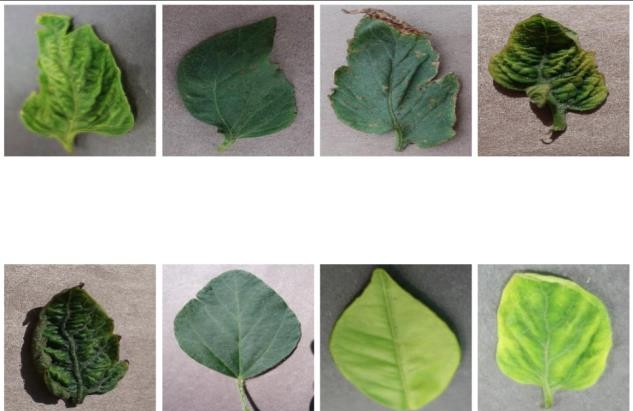}
  \caption{Dataset image samples}
\end{figure}

Further all the image pixels were normalized between 0 to 1 by division with 255 and split into their RGB channels after which GCC was calculated to be fed as a feature to the SVM layer. \textbf{Fig.5} Shows the RGB split of a sample image. Further analysis was done by comparing GCC values of healthy crops to the diseased crops and \textbf{Fig.6} compares the cumulative Green chromatic indices for healthy and diseased crops. The mean value of GCCs for healthy crops was 0.42 whereas for diseased crops was 0.34.
\begin{figure}[h]
  \centering
  \includegraphics[width=0.7\textwidth]{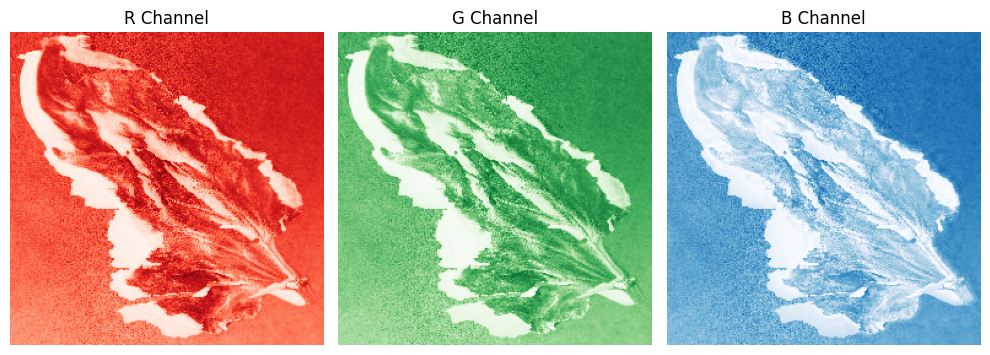}
  \caption{RGB Channel visualization for a sample leaf image}
\end{figure}
\\
The analysis on distribution as well as mean values indicated that GCC values of diseased crops were inherently less than that of healthy crops for that plant. 
\begin{figure}[h]
  \centering
  \includegraphics[width=0.6\textwidth]{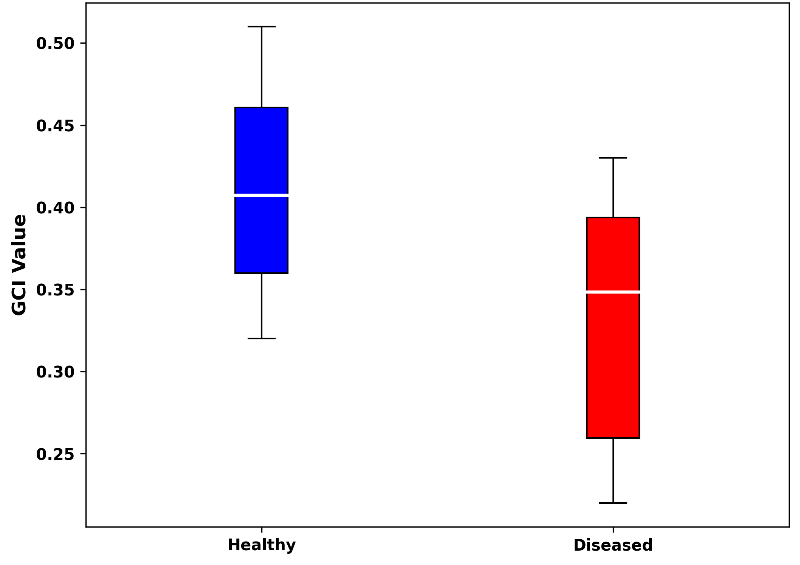}
  \caption{GCC Value Distribution Box-Plot: Diseased vs Healthy crop leaves}
\end{figure}
\\
The transformer backbone used in proposed approach uses the ViT-Base architecture. Larger ViT architectures could not be used due to hardware limitations. Image size was set at 256x256 for all the models. The activation function used for the ViT backbone was ‘SoftMax’ whereas all the subsequent dense layers used ‘Categorical Cross Entropy’ as the loss function with label smoothing set at 0.2. \textbf{Table.4} illustrates the architecture of the classification block over the ViT backbone. The VGG-16 and InceptionV3 model were implemented using the inbuilt application module in TensorFlow. Both the models used pretrained weights from the ImageNet Dataset. The output of these models was flattened into a 1-D Tensor using ‘Flatten’ layer and then ‘Dropout’ layer is added with a dropout rate set at 0.5. Finally, a dense layer with 39 neurons and SoftMax activation function is used for classification. ‘Adam’ optimizer is used as the optimization function for both the models, along with Categorical Cross Entropy loss. All the models were trained for a total of 50 epochs with batch size set at 32 and post analysis and performance comparison of model. \textbf{Table.3} gives an overview of the experimental parameters utilized for the experiment.\\

\begin{table}[h]
\caption{Experimental parameters}
\centering
\renewcommand{\arraystretch}{1.2} 
\begin{tabular}{ccc}
\hline
\ Function & Parameter & Value \\
\hline
\multirow{7}{*}{Image augmentation}
& Rotation by a random angle in degrees &	range [ -25 , 25 ]\\
& Random shifting across the width &	0.1\\
& Random shifting across the height &	0.1\\
& Random shearing &	0.2\\
& Horizontal and vertical flipping &	TRUE\\
& Random zooming &	0.2\\
& Rescale &	1./255\\
\hline
\multirow{2}{*}{Dataset Split}
& Training &	0.8\\
& Validating and Testing &	0.2\\
\hline
\multirow{4}{*}{Training parameters}
& Batch size &	32\\
& Epochs &	50\\
& Optimizer RMSprop &	adam\\
& Initial learning rate &	1.00e - 04\\    
\hline
\ Classifier (SVM)& Regularizer  & L2(0.01)\\
\hline
\ GCC  & Normalization of pixels  &	0-1\\
\hline
\multirow{4}{*}{ViT parameters}
& Patch size &	4\\
& Projection dimension &	64\\
& Number of heads &	4\\
& Transformer layers &	8\\    
\hline   
\end{tabular}
\end{table}

For analyzing the performance of the disease classification model 4 metrics namely Accuracy, Precision, Recall and F1-score have been utilized which are described as follows:

(a)	Precision

It measures the proportion of correctly predicted positive instances among the total predicted positive instances.
\begin{align}
\text{Precision} = \frac{TP}{TP + FP}
\end{align}

(b)	Accuracy

It measures the overall correctness of the model's predictions by comparing the number of correct predictions with the total number of predictions.
\begin{align}
\text{Accuracy} = \frac{TP + TN}{TP + TN + FP + FN}
\end{align}

(c)	Recall (Sensitivity or True Positive Rate)

It measures the proportion of correctly predicted positive instances among the total actual positive instances.
\begin{align}
\text{Recall} = \frac{TP}{TP + FN}
\end{align}

(d)	F1-Score

It is a harmonic mean of precision and recall, providing a balanced measure that considers both metrics.

\begin{align}
\text{F1 Score} = \frac{2 \times (\text{Precision} \times \text{Recall})}{\text{Precision} + \text{Recall}}
\end{align}
Where TP refers to True Positives (correctly predicted positive instances), TN refers to True Negatives (correctly predicted negative instances), FP refers to False Positives (incorrectly predicted positive instances), and FN refers to False Negatives (incorrectly predicted negative instances).

\begin{table}[h]
\caption{Model Summary over ViT based feature extraction}
\centering
\renewcommand{\arraystretch}{1.2} 
\small
\begin{tabularx} {0.7\textwidth}{ccc}
\hline
\ Layer &	Output Size &	Activation\\
\hline\hline
\ ViT-Base32 Output Layer ( Dense )&	768&	GeLU\\
\ Dense&	128&	ReLU\\
\ SVM&	39&	Softmax\\
\hline
\end{tabularx}
\end{table}

\section{Results}
Both the transfer learning models and the proposed approach were analyzed using multiple metrics \textbf{Table.5} shows the comparison between accuracy of the models used in this work.

\begin{table}[h]
\caption{Cross model performance comparison}
\centering
\renewcommand{\arraystretch}{1.2} 
\small
\begin{tabularx} {0.6\textwidth}{ccccc}
\hline
\ Model &	Accuracy &	Precision &	Recall &	F1-Score\\
\hline\hline
\ VGG-16 &	88.49 &	90.01 &	88.03 &	88.76\\
\ InceptionV3 &	79.93 &	81.88 &	76.43 &	76.21\\
\ GCCViT-SVM &	99.69 &	99.63 &	99.57 &	99.59\\
\hline
\end{tabularx}
\end{table}

VGG-16 model achieved an accuracy score of 88.49\%. Though, it is to be noted that VGG-16 model performs better over a higher number of epochs, hence, training the model for a greater number of epochs could result in considerably higher performance. However, this score still falls considerably short of the accuracy of our proposed approach. It should also be noted that the accuracy of the model varied considerably in each training epoch and the model loss on both the training and validation set was on an increasing trend throughout the training as shown in \textbf{Fig. 7(a)} and \textbf{7(b)}. The InceptionV3 model performed poorly, achieving an accuracy of only 79.93\%. It was also plagued by the same problem of an overall increasing loss during the training.
\\
\begin{figure}[h]
  \centering
  \includegraphics[width=\textwidth]{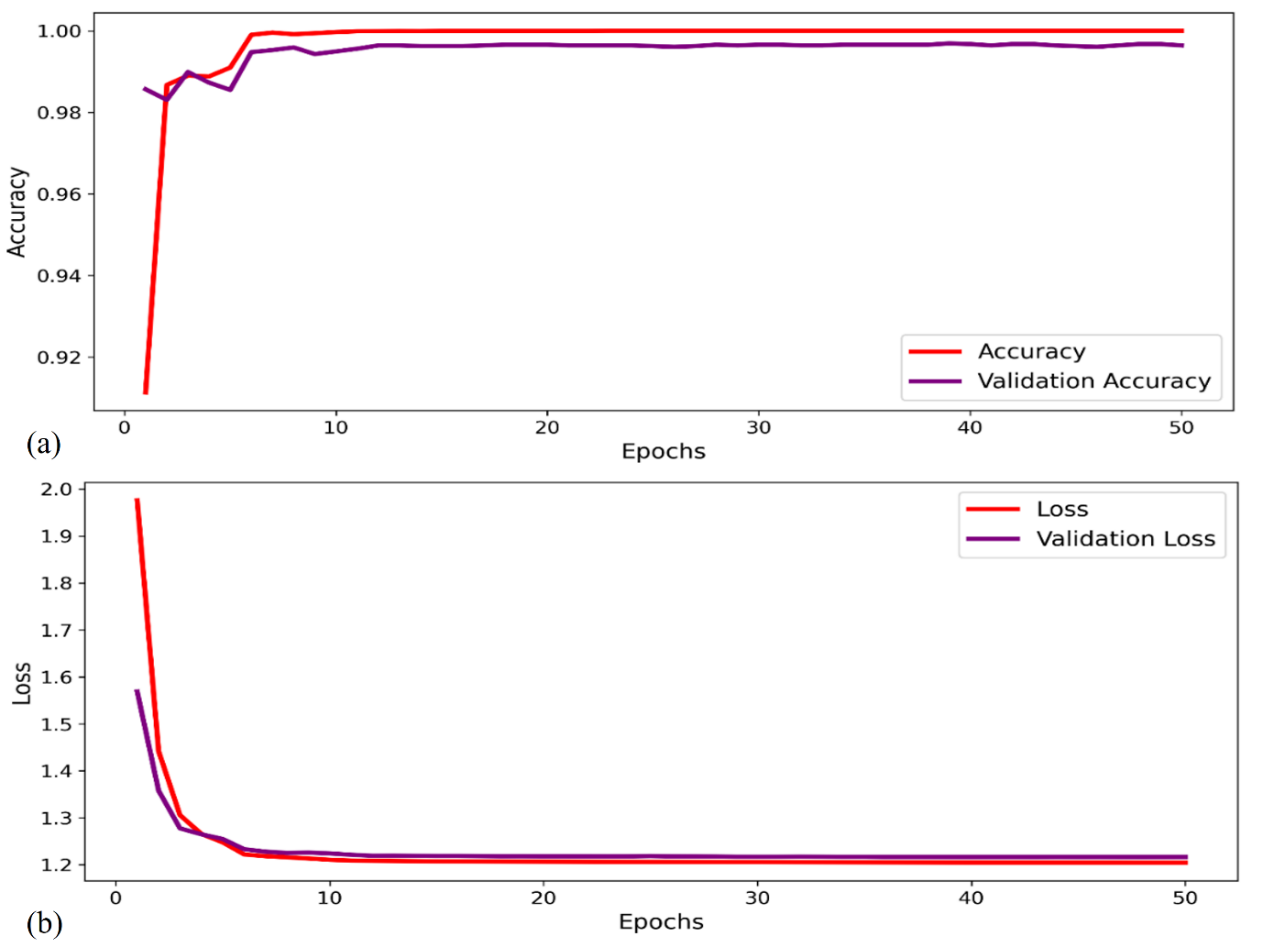}
  \caption{(a) GCCViT-SVM accuracy on training and validation set (b) GCCViT-SVM loss on training and validation set}
\end{figure}
\\
On comparison with recent work, results of this study indicate tremendous potential for the mentioned classification task, \textbf{Table 7} depicts the model performance comparison of our model with some of the recent studies. The proposed model achieved an accuracy of 99.69\% for the plant Village dataset which is higher than the work discussed in \textbf{Table 7}. It can be observed that attention based networks show better performance since they extract additional positional embeddings, but even in comparison to approaches based on vision transformer, the proposed model is less complex while still ensuring superior performance pertaining to the additional feature leveraged by calculating Green Chromatic Coordinate for input data. Also, it is to be noted that the model achieved higher accuracy in training and validation set much earlier in the training phase than the other two models. The model loss on both the training and test set was also decreasing throughout the training phase, and was considerably lower than the other models. 

In terms of class-wise performance, the proposed model achieved a perfect score across all 3 metrics for more than 15 classes. The class wise overview of model performance is given in\textbf{ Table.6}. Whereas \textbf{Fig.8} depicts the confusion matrix for the model across all 39 classes labelled 0 – 38 in the same order as in Table 6.

By quantizing the proposed model post training and converting both activations and weights to 8-bit precision, 4 times reduction in size of the model was observed with approximately only 2.28\% loss in accuracy. \textbf{Table.8} gives an overview of the changes post quantization on the proposed model.
\\
\begin{figure}[h]
  \centering
  \includegraphics[width=0.85\textwidth]{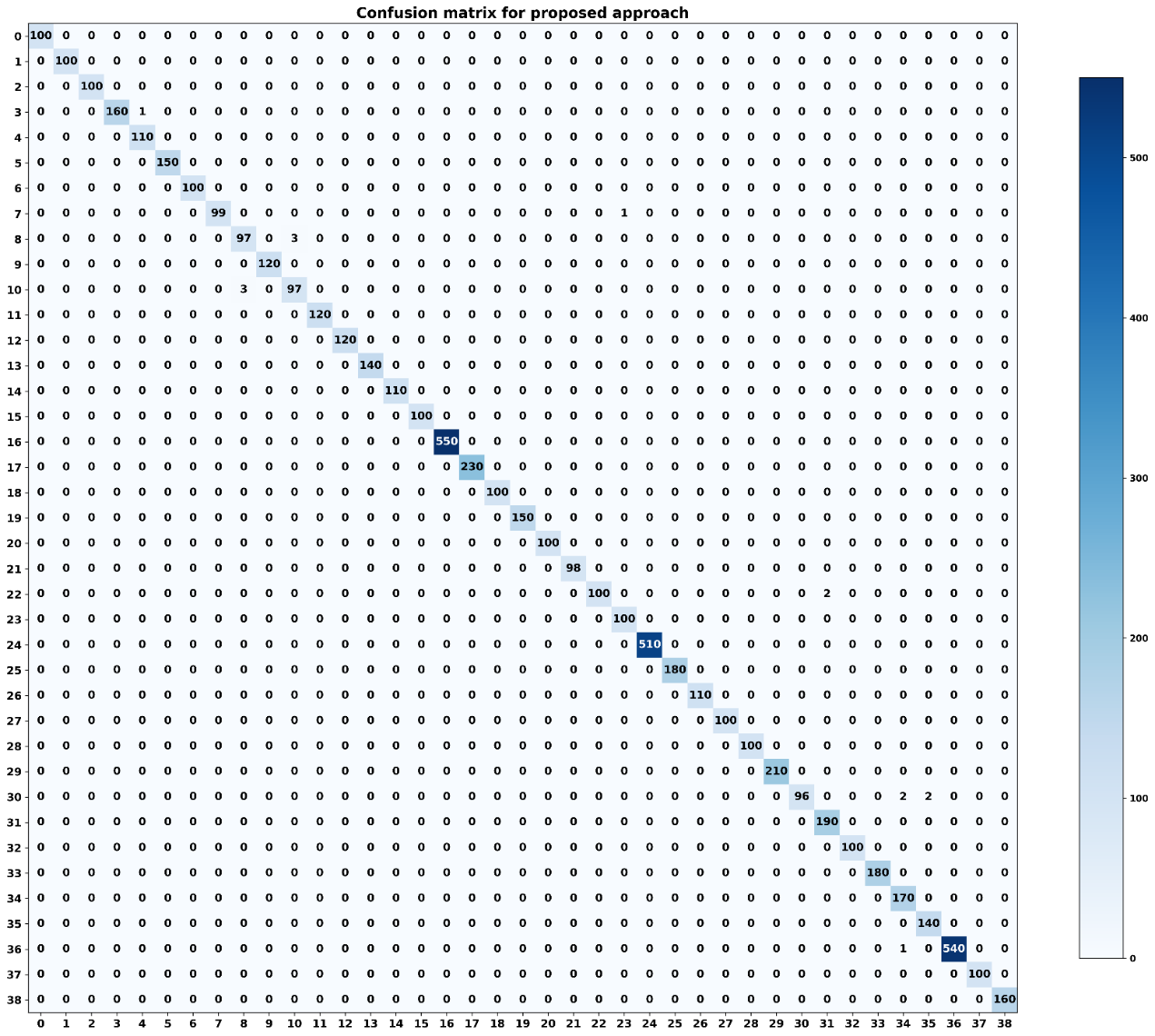}
  \caption{Confusion matrix for GCCViT-SVM}
\end{figure}
\\
\section{Discussions}
This section presents a comprehensive discussion on the findings and implications of the proposed methods. Key aspects of the study are addressed, including the challenges posed by the dataset, the nature of image processing and feature extraction, and experimental parameter optimization. Additionally, potential avenues have been discussed for future work and highlight the significance of the findings in the broader context of image classification.
\begin{table}[h]
\caption{Class-wise performance of GCCViT-SVM}
\centering
\small
\begin{tabularx} {0.8\textwidth}{ccccc}
    \hline
\ Label & 	Class & 	Precision & 	Recall & 	F-1 Score \\
    \hline\hline
\ 0 & 	Apple Scab  & 	1 & 	1 & 	1 \\
\ 1 & 	Apple Black Rot & 	1 & 	1 & 	1 \\
\ 2 & 	Apple Cedar apple rust & 	1 & 	1 & 	1 \\
\ 3 & 	Apple healthy & 	1 & 	1 & 	1 \\
\ 4 & 	Background without leaves & 	1 & 	0.99 & 	1 \\
\ 5 & 	Blueberry healthy & 	1 & 	1 & 	1 \\
\ 6 & 	Cherry healthy & 	1 & 	1 & 	1 \\
\ 7 & 	Cherry Powdery mildew & 	1 & 	0.99 & 	0.99 \\
\ 8 & 	Corn Cercospora leaf spot & 	0.99 & 	0.98 & 	0.98 \\
\ 9 & 	Corn Common rust & 	1 & 	1 & 	1 \\
\ 10 & 	Corn healthy & 	0.98 & 	0.99 & 	0.99 \\
\ 11 & 	Corn Northern Leaf Blight & 	1 & 	1 & 	1 \\
\ 12 & 	Grape Black rot & 	1 & 	0.99 & 	1 \\
\ 13 & 	Grape Esca (Black Measles) & 	0.99 & 	1 & 	1 \\
\ 14 & 	Grape healthy & 	0.99 & 	1 & 	1 \\
\ 15 & 	Grape Leaf blight (Isariopsis Leaf Spot) & 	1 & 	1 & 	1 \\
\ 16 & 	Orange Haunglongbing (Citrus greening) & 	1 & 	1 & 	1 \\
\ 17 & 	Peach Bacterial spot & 	1 & 	1 & 	1 \\
\ 18 & 	Peach healthy & 	1 & 	1 & 	1 \\
\ 19 & 	Bell pepper Bacterial spot & 	1 & 	1 & 	1 \\
\ 20 & 	Bell pepper healthy & 	1 & 	1 & 	1 \\
\ 21 & 	Potato Early blight & 	1 & 	1 & 	1 \\
\ 22 & 	Potato healthy & 	1 & 	1 & 	1 \\
\ 23 & 	Potato Late blight & 	1 & 	1 & 	1 \\
\ 24 & 	Raspberry healthy & 	1 & 	1 & 	1 \\
\ 25 & 	Soybean healthy & 	1 & 	1 & 	1 \\
\ 26 & 	Squash Powdery mildew & 	1 & 	1 & 	1 \\
\ 27 & 	Strawberry healthy & 	1 & 	1 & 	1 \\
\ 28 & 	Strawberry Leaf scorch & 	1 & 	1 & 	1 \\
\ 29 & 	Tomato Bacterial spot & 	1 & 	1 & 	1 \\
\ 30 & 	Tomato Early blight & 	0.99 & 	0.97 & 	0.98 \\
\ 31 & 	Tomato healthy & 	1 & 	0.99 & 	1 \\
\ 32 & 	Tomato Late blight & 	0.99 & 	1 & 	1 \\
\ 33 & 	Tomato Leaf Mold & 	1 & 	1 & 	1 \\
\ 34 & 	Tomato Septoria leaf spot & 	0.98 & 	0.99 & 	0.99 \\
\ 35 & 	Tomato Spider mites & 	0.98 & 	0.99 & 	0.99 \\
\ 36 & 	Tomato Target Spot & 	1 & 	1 & 	1 \\
\ 37 & 	Tomato mosaic virus & 	1 & 	1 & 	1 \\
\ 38 & 	Tomato Yellow Leaf Curl Virus & 	1 & 	0.99 & 	1 \\
\hline
\end{tabularx}
\end{table}
\
\\
\begin{table}[h]
\caption{Performance Comparison with recent work on Plant Village}
\centering
\small
\small
\renewcommand{\arraystretch}{1.5} 
\begin{tabularx}{\textwidth}{cp{3cm}ccccc}
\hline
\multicolumn{1}{c}{\makecell{Referred studies}}  & \multicolumn{1}{c}{\makecell{Algorithm (classification)}}& \multicolumn{1}{c}{\makecell{Accuracy}} & \multicolumn{1}{c}{\makecell{Precision}}& \multicolumn{1}{c}{\makecell{Recall}}& \multicolumn{1}{c}{\makecell{F1-Score}}\\
\hline\hline
\ Tabbakh and Barpanda, 2022\cite{R9}  & 	Machine learning model using LGBM model & 	0.9439 & 	0.9475 & 	0.9476 & 	0.9472 \\
\  Mohana et al., 2021 \cite{R11}& 	CNN & 	0.9677 & 	0.9646 & 	0.9625 & 	0.9635 \\
\ Sachdeva et al., 2021\cite{R12} & 	Deep Convolutional neural network with Bayesian learning & 	0.989 & 	0.982 & 	0.979 & 	0.9804 \\
\ Jasim and Al-Tuwaijari, 2020\cite{R13} & 	CNN & 	0.9803 & 	0.9827 & 	0.9801 & 	0.9814 \\
\ Shijie, et al.,  2017 \cite{R14}& 	VGG16 & 	0.88 & 	- & 	- & 	- \\
\ Karthik et al., 2020\cite{R30} & 	Attention and the residual network & 	0.9583 & 	0.962 & 	0.956 & 	0.9589 \\
\ Chen et al.,  2021a \cite{R19}& 	DenseNet along with attention approach & 	0.9794 & 	0.8959 & 	0.8671 & 	0.8807 \\
\ Chen et al., 2021b \cite{R20}& 	pre-trained MobileNet-V2 and attention mechanism & 	0.9668 & 	0.9749 & 	0.9583 & 	0.9664 \\
\ Thakur et al., 2021 \cite{R21}& 	PlantViT & 	0.9861 & 	0.9824 & 	0.9833 & 	0.9828 \\
\ Tabbakh and Barpanda, 2023\cite{R22} & 	TLMViT (VGG19 followed by ViT ) & 	0.9881 & 	0.9872 & 	0.9876 & 	0.9873 \\
\ Proposed model & 	GCCViT-SVM (GCC and ViT output features fed to SVM ) & 	0.9969 & 	0.9978 & 	0.9965 & 	0.9971 \\
\hline
\end{tabularx}
\end{table}
\\
\subsection{Imabalanced Dataset}
One of the challenges encountered in this study was the imbalanced nature of the dataset. The dataset consisted of 39 classes, but the number of samples per class varied significantly. This class imbalance poses a significant obstacle for image classification tasks as it can lead to biased model performance. To address this issue, we employed techniques such as data augmentation and stratification of split during training. Data augmentation techniques, including rotation, flipping, and scaling, were used to generate additional samples for underrepresented classes. Moreover, the use of stratified split provided a way to assign same ratio of test and train sets amongst the minority as well as majority classes, thereby reducing the impact of the class imbalance on the model's performance. However, it must be noted that better balanced datasets can influence the performance of the model positively and make it further scalable.
\\
\begin{table}[h]
\caption{Model Comparison before and after quantization}
\centering
\renewcommand{\arraystretch}{1.2} 
\begin{tabularx}{0.9\textwidth}{ccc}
\hline
Model & Before Quantization & After Quantization (TF-Lite format) \\
\hline\hline
Accuracy & 99.69\% & 97.41\% \\
Size & 338.4 MB & 84.6 MB \\
Weights and Activations & 32 bit & 8 bit \\
\hline
\end{tabularx}
\end{table}
\\

\subsection{Image processing and feature extraction}
The Images utilized for the experiment, when fed to the model rely completely on the model’s capability to extract features. To ensure the integration of relevant techniques which align with the agricultural literature, Indices like Green Chromatic Coordinate was utilized to analyze the crop health and the spread of the disease. 

It must be noted that further experiments pertinent to agricultural methodologies, such as analysis of near-infrared reflectance, are required to analyze the density of green in the crops or vegetation, which can further enhance the model's capability in detecting diseases and improve large-scale implementation and usability.\\
\subsection{Hyperparameters}
Optimizing hyperparameters is a critical step in achieving optimal model performance. This study implemented various experimental parameter tuning approaches to identify the best configuration for our image classification model. Grid search and random search were utilized to explore a wide range of parameter combinations and evaluate their impact on the model's performance. Additionally, techniques such as cross-validation to estimate the generalization performance of different hyperparameter settings and ensure the reliability of the results were leveraged. In terms of reducing the memory footprint and ensuring low-computational resource device compatibility (IoT, Mobiles etc.) Post-training quantization was used. Further fine-tuning of hyperparameters, including learning rate, batch size, quantization aware training, mixed quantization, and regularization strength, may significantly improve the model's ability to generalize to unseen data, increase overall classification accuracy, and reduce memory footprint.
\subsection{Deployment and user interfaces }
Farmers and agriculture have historically served as the cornerstone of the Indian economy. This research not only presents its findings but also lays the groundwork for the development of a scalable and cost-effective solution that can be readily implemented to benefit the agricultural community. The comprehensive website, built upon this model to maximize accuracy, is currently in the developmental phase. Future work will encompass the creation of a mobile application and the integration of additional algorithms for crop recommendation and soil nutrition monitoring, aiming to create an all-encompassing solution.

\section{Conclusions}
In this paper, a hybrid model called GCCViT-SVM is proposed for plant disease classification, combining Support Vector Machines with a vision transformer. The GCCViT-SVM model consists of four stages: data acquisition, image augmentation, feature extraction with GCC calculation, and classification. Initially, the model is trained and evaluated using the PlantVillage dataset and all its 39 classes. Image augmentation techniques, including rotation, shifting, shearing, zooming, and flipping, are applied to increase the number of training samples. The feature extraction process involves a vision transformer that is employed to extract deep features based on the initial features and then GCC for each image is calculated. Finally, a MLP head classifier along with SVM determines the class to which the leaf belongs. The performance of the proposed GCCViT-SVM model is evaluated using various metrics such as accuracy, loss, precision, F1-score, and recall. Experimental results show that the model achieves high accuracy, and the validation accuracy reaches 99.69\% with corresponding training accuracy of 100\%. Furthermore, the performance of GCCViT-SVM is compared to transfer learning-based models (demonstrating the effectiveness of incorporating attention based networks). GCCViT -SVM outperforms the transfer learning-based model, achieving higher validation accuracy and lower validation loss for all classes of the dataset. It outperforms VGG -16 by 11.2\% and inception v3 by 19.7\% in terms of accuracy. The findings of this study highlight the benefits of data augmentation, the effectiveness of ViT for deep feature extraction with traditional remote sensing and image processing based GCC, and classification capability of traditional Machine learning classification algorithms, SVM for the scope of this experimentation. The model is then converted into TF. Lite model includes size reduction by quantization to ensure compatibility with IoT devices, which can then be accessed through mobile or web-app, achieving a single stop software solution for crop disease classification. Further research is encouraged with improvements in dataset quality, integration, and experimentation with multiple image processing and feature extraction techniques for better performance and stability. 

\bibliographystyle{unsrt}
\bibliography{ref}

\end{document}